\documentclass[aps,prl,twocolumn,superscriptaddress,showpacs,showkeys]{revtex4-1}

\pdfoutput=1

\usepackage[utf8]{inputenc}%
\usepackage{lmodern}%
\usepackage[T1]{fontenc}%
\usepackage{hyperref}%
\usepackage{graphicx}%
\usepackage{amsmath,amssymb}%

\DeclareMathOperator{\argmax}{arg \, max}%

\begin{document}

\title{Depinning as a coagulation process}

\author{Melih \.{I}\c{s}eri}%
\email{melih.iseri@boun.edu.tr}%
\affiliation{Physics Department, Bo\u{g}azi\c{c}i University, Bebek
  34342 Istanbul, Turkey}%

\author{David C. Kaspar}%
\email{david\_kaspar@brown.edu}%
\thanks{supported by NSF DMS-1148284}%
\affiliation{Division of Applied Mathematics, Brown University,
  Providence, RI 02912, USA}%

\author{Muhittin Mungan}%
\email{mmungan@boun.edu.tr}%
\thanks{supported by grant 14B03P6 of Bo\u{g}azi\c{c}i University}%
\affiliation{Physics Department, Bo\u{g}azi\c{c}i University, Bebek
  34342 Istanbul, Turkey}%

\date{\today}

\begin{abstract}
  We consider a one-dimensional sandpile model which mimics an elastic
  string of particles driven through a strongly pinning periodic
  environment with phase disorder.  The evolution towards depinning
  occurs by the triggering of avalanches in regions of activity which
  are at first isolated but later grow and merge.  For large system
  sizes the dynamically critical behavior is dominated by the
  coagulation of these active regions. Our analysis of the evolution
  and numerical simulations show that the observed sizes of active
  regions is well-described by a Smoluchowski coagulation equation,
  allowing us to predict correlation lengths and avalanche sizes.
\end{abstract}

% PACS, the Physics and Astronomy Classification Scheme.  Insert
% suggested PACS numbers in braces on next line.
% 05.40.-a Fluctuation phenomena, random processes, noise, and
% Brownian motion
%
% 64.60.Ht Dynamic critical phenomena
%
% 45.70.Ht Avalanches
%
% 46.65.+g Random phenomena and media

% Use showkeys class option if keyword display desired.

\maketitle

\paragraph{Introduction ---}

Chains of particles connected by springs, where each particle
experiences a randomly shifted periodic potential and an external
driving force, have served as phenomenological models for charge
density waves (CDWs) \cite{Gruner,FuLee}.  Of particular interest is
the \emph{depinning} transition as the external force increases to a
threshold value beyond which the system slides, which is considered a
\emph{dynamic critical phenomenon}
\cite{DSFisher83,DSFisher85,Littlewood86,Middleton93,MyersSethna93,snc90,snc91}.
The behavior of such systems has been of interest in diverse areas,
such as flux lines in type II superconductors \cite{Blatter}, fluid
invasion in porous media \cite{Wilkinson}, propagation of cracks
\cite{Bouchaud,Alava}, friction and earthquakes \cite{Kawamura12}, and
plastic flows in solids, where dislocational structures depin under
shear load \cite{Salman11,Salman12}.  These systems being far from
equilibrium, the mechanisms leading to critical behavior and universal
features are not yet well-understood.  This is due, at least in part,
to a lack of simple models admitting detailed analysis.

The articles \cite{KM13,KM14} introduced one such model, a
one-dimensional sandpile emerging in the strong pinning limit of a CDW
system which may be driven along a transverse axis.  When the external
force varies slowly compared to the relaxation times, the evolution to
threshold occurs by avalanches corresponding to the local depinning of
segments.  The resulting critical behavior is sensitive to initial
conditions (ICs): starting from the final static configuration
obtained driving in the $(-)$ direction, and then driving in the $(+)$
direction, the evolution towards threshold can be described explicitly
and proceeds by the growth of a \emph{single} depinned segment
arrested only by its endpoints.  The situation is markedly different
when one considers generic ICs: the evolution proceeds by
\emph{multiple} depinned segments that each grow and
\emph{merge}. Numerics indicate \cite{KM13,KM14} that there is a
critical transition, with different scaling behavior, agreeing with
the predictions of Narayan {\em et al.}
\cite{NarayanMiddleton94,NarayanFisher92}.  In this letter we report
extensive simulation results concerning the evolution starting from a
macroscopically flat initial condition.  Our analysis reveals an
unexpectedly clean connection with mean-field coagulation phenomena,
providing a new viewpoint on macroscopic features of the depinning
transition and shedding light on the emergence of universality in
dynamical critical phenomena \footnote{See \cite{FournierBressaud09}
  for a one-dimensional model which also involves avalanches and a
  kinetic equation, and the recent work \cite{Beznea2016} which
  involves avalanches and fragmentation.  These models are otherwise
  quite different from ours and seem unrelated to
  depinning.}\nocite{FournierBressaud09,Beznea2016}.

\paragraph{The model ---}

We begin by recalling the toy model of \cite{KM13,KM14}.  Fix a large
integer $L$ and consider $L$-periodic vectors $\vec{z}, \vec{m}$, and
$\vec{\rho}$, related as follows:
\begin{equation}
  \label{eq:zvsm}
  z_i = \rho_i + \Delta m_i = \rho_i + m_{i-1} - 2 m_i + m_{i+1}.
\end{equation}
Here $\vec{\rho}$ represents the quenched phase disorder, $\vec{m}$
counts the number of potential wells through which the particles are
displaced, and $\vec{z}$ corresponds to a suitable rescaling of the
displacements of the particles from the centers of their wells.
Unlike the standard sandpile \cite{BTW87,Dhar,Redig05}, heights $z_i$
have fractional parts from $\rho_i$ which persist, since the $m_i$
take only integer values, and the dynamics are deterministic and
extremal \cite{Paczuski96}. The CDW process of raising the force until
a particle crosses wells and waiting for the system to relax to a new
static configuration is equivalent \cite{KM14} to applying the
following \emph{avalanche algorithm}:
\begin{itemize}
\item[A1.] Record the critical height $h_c = \max_i z_i$.
\item[A2.] While there exists $i$ such that $z_i \geq h_c$, replace
  \begin{equation}
    \begin{gathered}
    m_i \to m_i + 1, \,
    z_i \to z_i - 2, \, z_{i\pm 1} \to z_{i\pm 1} + 1,
    \end{gathered}
  \end{equation}
  repeating as necessary until $\max_i z_i < h_c$.
\end{itemize}
This is precisely sandpile \cite{Dhar} toppling at critical height
$h_c$, and a standard argument \cite{Redig05} can be adapted to show
that the result is independent of the order in which the indices $i$
are chosen in step A2.  Furthermore, it can be shown \cite{KM14} that
there is a unique number $z^+_{\max}$, the threshold height, such that
the above algorithm terminates if and only if $h_c > z^+_{\max}$.

\begin{figure*}
  \mbox{\includegraphics{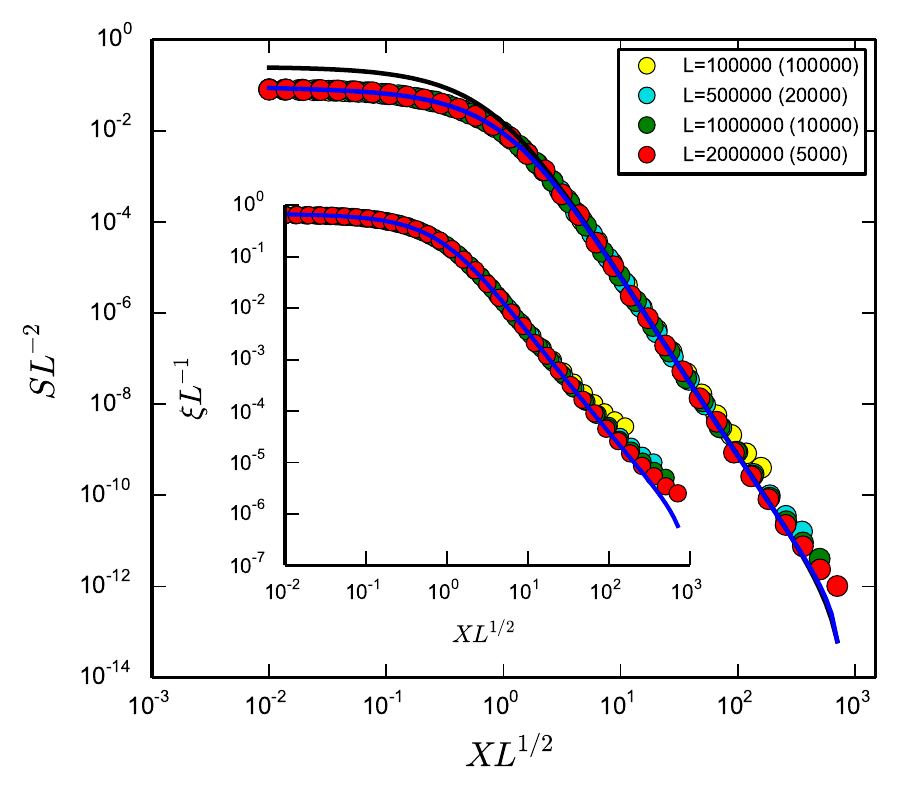} \hspace{-3ex}
    \includegraphics{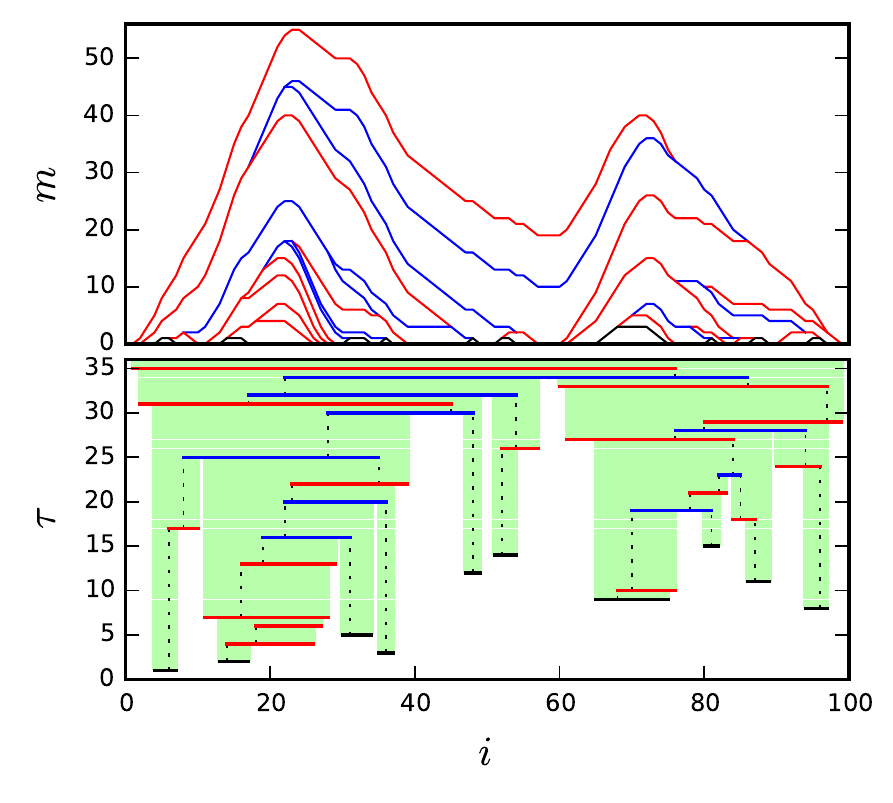}}
  \caption{\label{fig:plot-av-ar} Left: finite-size scaling for
    avalanche size and correlation length (inset), including
    simulation data (circles) and predictions (curves; see
    \eqref{eq:xi-s-moments}) for different system sizes $L$.
    Parenthesized numbers in the legend give the number of independent
    random realizations of the disorder.  Top right: the evolution of
    well numbers $\vec{m}$ via avalanches for a small system of size
    $L = 100$.  Bottom right: the corresponding avalanche intervals
    and partitions, illustrating the active regions (ARs) of size
    $> 1$ in light green.  Both right: avalanches shown in black, red,
    and blue involve respectively 0, 1, and $\geq 2$ non-singleton
    ARs.}
\end{figure*}

Define $i_{\max} = \argmax_i z_i$, when this is unique \footnote{This
  always holds in the case of absolutely continuous disorder
  $\vec{\rho}$}.  Upon termination of the algorithm, $\vec{z}$ has
changed by
\begin{equation}
  \label{eq:zupdate}
  \begin{gathered}
    \begin{aligned}
      z_{i_{\max}} &\to z_{i_{\max}} - 1, & z_{i_m} &\to z_{i_m} - 1 \\
      z_{i_l} &\to z_{i_l} + 1, & z_{i_r} &\to z_{i_r} + 1
    \end{aligned}
    \\
    i_m = i_r + i_l - i_{\max}
  \end{gathered}
\end{equation}
where $i_l,i_r$ are the first sites to the left and right of
$i_{\max}$ satisfying $z_i < h_c - 1$, and $i_m$ is the reflection
\footnote{When $i_{\max} = i_m$, the change is
  $z_{i_{\max}} \to z_{i_{\max}} - 2$.} of $i_{\max}$ across the
midpoint of the interval $i_l,i_r$ \cite{KM14}.  (Above and henceforth
all addition and subtraction of indices is to be understood modulo
$L$, with results in $[0,L)$.)  The change in $\Delta \vec{m}$ is the
change in $\vec{z}$, so $\vec{m}$ is modified by adding a nonnegative
trapezoidal bump with slopes $\{0,\pm 1\}$ and corners at
$i_l,i_{\max},i_m,i_r$.  We will refer to the periodic interval
$[i_l, i_r]$ as an avalanche segment. The length of the segment
$i_r - i_l + 1$ furnishes a correlation length $\xi$.  Defining the
size $\mathcal{S}$ of the avalanche as the total number of well jumps
that occurred, it can be shown \cite{KM14} that
$\mathcal{S} = (i_{\max} - i_l)(i_r - i_{\max})$.

Observe that $z_{\max}$ decreases under repeated application of the
algorithm.  Let $\tau = 0,1,2,\ldots$ index the observed
configurations after $\tau$ \emph{complete} executions of the
algorithm, and define the control parameter $X$ as
\begin{equation}
  \label{eq:X}
  X(\tau) = z_{\max}(\tau) - z^+_{\max} \geq 0.
\end{equation}
We have $X(\tau) = 0$ if, after $\tau$ avalanches, we reach the
(essentially unique) \emph{threshold configuration} with
$z_{\max} = z_{\max}^+$, which gives the final shape of the chain
prior to complete depinning.  For the toy model this configuration can
be explicitly constructed, yielding both a scaling limit for the
threshold well numbers $\vec{m}^+$ as $L \to \infty$ and a
characterization of the threshold force \cite{KM14}.

Here we discuss the case of a flat initial configuration,
$\vec{m}(0) = \vec{0}$, with disorder $\rho_i$ \emph{i.i.d.}\ uniform
on $[-1,+1]$, where the evolution to threshold is illustrated in
Figure~\ref{fig:plot-av-ar}.  The left panel shows the finite-size
scaling behavior of the expected avalanche size $\mathcal{S}$ and
correlation length $\xi$ obtained from extensive simulations,
indicating that in the limit of large $L$, the depining transition is
critical:
\begin{equation}
  \label{eq:scalinglaw}
  \mathcal{S} \sim X^{-\gamma} %
  \quad \text{and} \quad %
  \xi \sim X^{-\nu} %
  \quad \text{with} \quad %
  \gamma = 4, \, \nu = 2,
\end{equation}
as predicted in \cite{NarayanMiddleton94,NarayanFisher92}.  The two
panels on the right ilustrate the microscopic details of the
evolution.  

\paragraph{Coagulation ---}

Under repeated application of the avalanche algorithm, the flat
initial configuration deforms via the depining of segments which are
initially separated, but grow and merge.  To understand this merging
process, we define after each iteration of the algorithm a set of
\emph{Active Regions} (ARs) $\Pi(\tau)$.  Each region is a periodic
interval in $\{0,\ldots,L-1\}$, and we denote
$[a,b] = \{a,a+1,\ldots,b\}$.  Initially we have $\Pi(0)$ equal to the
set of singleton intervals.  The $\tau^{\mathrm{th}}$ avalanche occurs
in some interval $[i_l,i_r]$, and we define $\Pi(\tau)$ to be the
finest partition of $\{0,\ldots,L-1\}$ which is coarser than the cover
$\Pi(\tau-1) \cup \{[i_l,i_r]\}$.  More intuitively, the new partition
joins together all those elements of the old partition which overlap
with $[i_l, i_r]$.  Note that the set of endpoints of intervals in
$\Pi(\tau)$ is exactly $\{ i : m_i = 0 \}$. The ARs and their
evolution are depicted by the light green shaded areas in 
Figure~\ref{fig:plot-av-ar} (bottom right).

In associating with the dynamics the partitions $\Pi(\tau)$ as defined
above, it seems that we have \emph{imposed} coagulation on the
problem, rasing the concern that the setup is contrived to yield the
desired result.  We emphasize the following Points:
\begin{itemize}
\item[P1.] We do not require the coagulation to be binary, and yet
  will find that binary events are macroscopically dominant over a
  large portion of the evolution.
\item[P2.] The relative rates of coagulation events are not evident
  in the setup but rather will emerge, in one of the nicest possible
  forms, in both the numerics and a heuristic calculation.
\item[P3.] Each AR has at most one \emph{stop site} which is pinned
  strongly enough to arrest avalanches.
\item[P4.] Since the avalanches which have occurred within the
  various active regions $\Pi(\tau)$ up to time $\tau$ have not
  interacted across the boundaries of $\Pi(\tau)$, we see that
  \emph{conditionally given $\Pi(\tau)$} we have
  \begin{equation}
    (z_i, m_i : i \in [a,b]), \quad [a,b] \in \Pi(\tau),
  \end{equation}
  statistically independent with distributions depending only on the
  sequence length $\ell = b - a + 1$.
\end{itemize}
We proceed to explain P3 in detail.  Suppose we have determined $h_c$
and completed an avalanche, resulting in the changes
\eqref{eq:zupdate}.  The new configuration has $\vec{z}$ satisfying
\begin{equation}
  \label{eq:AR-z-ineqs}
  h_c - 2 < z_{i_m} < h_c - 1 < z_i < h_c, \quad \forall i \in [i_l,i_r]
  \setminus \{i_m\}.
\end{equation}
The first two inequalities hold because $i_m$ either did not initiate
the avalanche and received $-1$, or $i_m = i_{\max}$ did initiate but
received $-2$.  The third holds because sites strictly between $i_l$
and $i_r$ were not capable of arresting the avalanche, and the sites
$i_l$ and $i_r$ were capable but each received $+1$.  We call $i_m$ a
stop site because this is capable of halting (one side of) a
subsequent avalanche avalanche above critical heights $h_c - \epsilon$
for some $\epsilon > 0$.  Stop sites may \emph{expire}: before having
the opportunity to stop an avalanche, the critical height might have
decreased more than $\epsilon$.  When an avalanche joins two or more
ARs, $i_l$ and $i_r$ must land on unexpired stop sites, \emph{using}
them by adding $+1$ and creating a single new stop site inbetween.  By
induction each active region has at most one stop site.  Note that by
establishing \eqref{eq:AR-z-ineqs}, an avalanche conditions an AR's
response to future avalanches.  This response is markedly different in
pristine areas where no avalanches have yet occurred: after one side
of an avalanche enters such an area, it will continue for a number of
sites whose average admits a small $O(1)$ upper bound (independent of
the system size) which holds uniformly over the whole evolution.

\paragraph{Mean-field statistics ---}

Our main observation is that the length statistics of the ARs recorded
in $\Pi(\tau)$ are numerically quite close to a well-known
\cite{Golovin63,Aldous} exact solution
\begin{equation}
  \label{eq:golovin}
  f(t, \ell) = e^{-t} B(1 - e^{-t}, \ell), \quad B(\lambda, \ell) =
  \frac{(\lambda \ell)^{\ell-1} e^{-\lambda \ell}}{\ell!},
\end{equation}
to the Smoluchowski coagulation equation with additive collision
kernel \cite{Smoluchowski16,Aldous,Norris99}:
\begin{multline}
  \label{eq:smoluchowski}
  \partial_t f(t,\ell) = \sum_{\ell' = 1}^{\ell-1} \frac{1}{2}
  \alpha(\ell', \ell - \ell') f(t,\ell') f(t,\ell-\ell') \\
  - \sum_{\ell' = 1}^\infty \alpha(\ell, \ell') f(t,\ell) f(t,\ell'),
\end{multline}
for $\ell = 1, 2, \ldots$, and in the additive case
$\alpha(\ell,\ell') = \ell + \ell'$.  This infinite-dimensional ODE
system describes binary aggregation in the mean-field setting:
$f(t,\ell)$ gives the number density per unit volume of clusters of
size $\ell$ at time $t$, where clusters of sizes $\ell$ and $\ell'$
interact to form a new cluster of size $\ell + \ell'$ at rate
$\alpha(\ell,\ell')$.  Equation \eqref{eq:smoluchowski}, with various
kernels $\alpha(\cdot,\cdot)$, has been used in modeling aerosols
\cite{Drake72}, formation of large scale structure in astronomy
\cite{SilkWhite78}, and aggregation of algae cells
\cite{AcklehFitzpatrickHallam94}.  

\begin{figure}
  \includegraphics{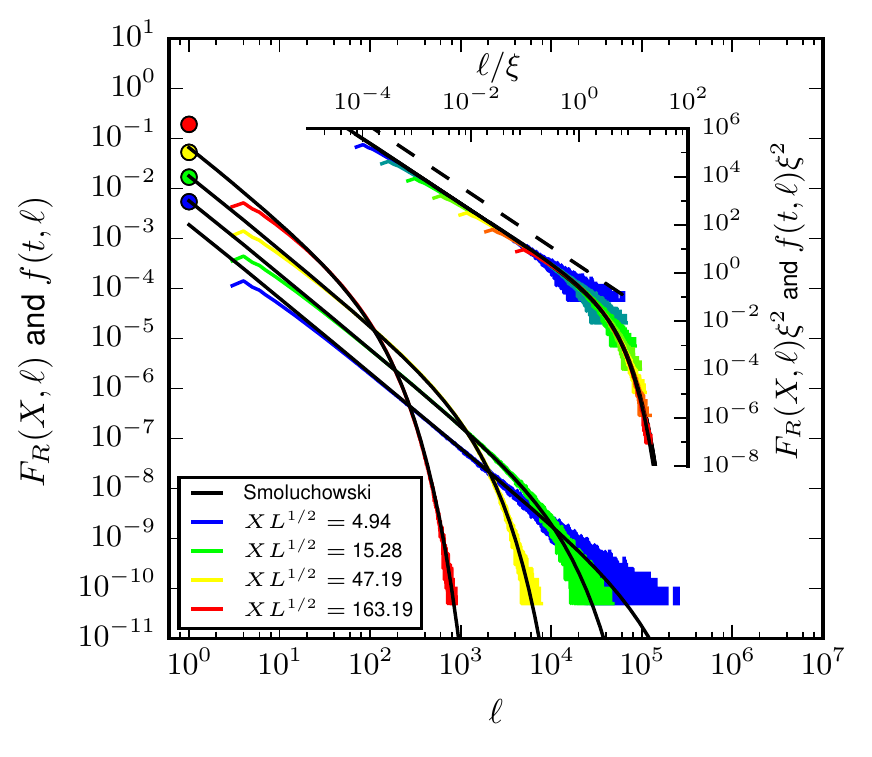}
  \caption{\label{fig:smol-dist} The empirical length distributions
    $F_R(X,\ell)$ (various colors) compared with the exact solution
    $f(t,\ell)$ to Smoluchowski \eqref{eq:golovin}, matching $X$ and
    $t$ by equating second moments.  In the inset, the axes are scaled
    so that the these collapse; the dashed line is a power law with
    exponent $-3/2$ and follows from \eqref{eq:golovin}. The agreement
    of distributions deteriorates prior to and after the scaling
    region visible in Figure~\ref{fig:plot-av-ar}; \emph{see} also the
    animation,
    %
    %in the Supplementary Material at [URL INSERTED BY PUBLISHER],
    available online \cite{Iseri16},
    which provides a dynamic view of the above.  The colored markers
    placed at $\ell = 1$ and the gap in the data between $\ell = 1$
    and $\ell = 3$ are artifacts of the definition of the partition
    $\Pi$: we cannot produce any ARs of length 2.}
\end{figure}

Associated with a realization of the toy model and its partitions
$\Pi(\tau)$ of ARs we have a size distribution
\begin{equation}
  N(\tau, \ell) = \sum_{[a,b] \in \Pi(\tau)} \delta_{b - a + 1,\ell},
\end{equation}
normalized so that $L^{-1} N(0, \ell) = f(0, \ell) = \delta_{\ell,1}$,
using Kronecker $\delta$ notation.  We give numerical evidence that
$f$ approximates a law of large numbers for $N$ at fixed times $\tau$.
For this we take many realizations, which we synchronize in time not
by the number of steps $\tau$ but rather $X(\tau)$ defined in
\eqref{eq:X}, which is the natural control parameter.  Given a
realization and $x \in [0,1]$, write $\vec{z}(x), \vec{m}(x)$ for the
\emph{first} configuration we observe with $X \leq x$, and likewise
write $\Pi(x)$ for its associated partition into ARs.  Given $R$
independent realizations with corresponding size distributions
$N_1(x,\ell),\ldots,N_R(x,\ell)$, define
\begin{equation}
  F_R(x, \ell) = (LR)^{-1} \sum_{k=1}^R N_k(x, \ell).
\end{equation}
We obtain $F_R$ via simulation with a Python library we have developed
for the toy model \cite{kmtoy16}.  After matching time scales $x$ for
$F_R$ and $t$ for $f$ by equating second moments,
\begin{equation}
  \sum_\ell \ell^2 F_R(x,\ell) = \sum_\ell \ell^2 f(t,\ell) = e^{2t},
\end{equation}
we plot in Figure~\ref{fig:smol-dist} distributions $f(t,\ell)$ and
$F_R(x,\ell)$ at various times.

\paragraph{Similarities explained ---}

We do not prove the law of large numbers suggested above, and indeed
do not claim that this is exactly given by $f$, but can offer a
heuristic Explanation of the similarities:
\begin{itemize}
\item[E1.] The probability that the next avalanche begins with a site
  inside an AR of length $\ell$ is similar to $\ell/L$.
\item[E2.] Since large ARs will initiate avalanches most often, their
  stop sites will tend to be used rather than expire, and the
  avalanches mostly extend in only one direction.
\item[E3.] On the side where the avalanche exits the triggering AR,
  it is very likely to stop if it hits another macroscopically-sized
  AR.  We thus expect to join to the triggering AR some small number
  of tiny ARs, which is not macroscopically observable, and (probably)
  at most one macrosocopic AR.
\end{itemize}
Combining E2 and E3, we expect that assuming binary coagulation yields
a reasonable approximation when we care primarily about large ARs, as
we will for the correlation length and avalanche size that we discuss
shortly.
\begin{itemize}
\item[E4.] Though the model is spatially ordered,
  \emph{statistically} it behaves as if it is well mixed.  In
  particular the length of the second AR in the avalanche is selected
  uniformly from the list of all AR lengths (excluding the triggering
  AR).  We explain further below.
\end{itemize}
Assuming that E1--E4 hold and that the system is large enough that
$N(\tau, \ell)$ is effectively deterministic, the expected change
$N(\tau+1, \ell) - N(\tau,\ell)$ is approximated by
\begin{equation}
  \label{eq:smoluchowski-emerges}
  \sum_{\ell' = 1}^{\ell-1} \frac{\ell' N(\tau, \ell')}{L}
  \frac{N(\tau, \ell - \ell')}{M_0(\tau)} - \frac{\ell N(\tau,
    \ell)}{L} - \frac{N(\tau, \ell)}{M_0(\tau)}, 
\end{equation}
having written 
\begin{equation}
  M_k(\tau) = \sum_{\ell = 1}^L \ell^k N(\tau, \ell)
\end{equation}
for the $k^{\mathrm{th}}$ moment of $N$.  The summation in
\eqref{eq:smoluchowski-emerges} is over those sizes which sum to
$\ell$, and reflects choosing a triggering AR with probability like
$\ell'/L$ and then a second AR uniformly from those which remain.  The
loss terms correspond to selection as a triggering AR or as a
secondary AR.  Symmetrizing the summation of
\eqref{eq:smoluchowski-emerges} in the variables $\ell', \ell-\ell'$
and factoring $L M_0(\tau)$ yields an equation matching
\eqref{eq:smoluchowski}, up to a change in time scale.

Regarding E4, recall that the Smoluchowski equation arises as a law of
large numbers for the Marcus-Lushnikov \cite{Marcus68,Lushnikov78}
stochastic coalescent as the number of clusters tends to infinity
\cite{Aldous,Norris99,Rezakhanlou13,FournierGiet04,FournierCepeda11}.
These models are well-mixed in the sense that any pair of clusters may
interact, which contrasts with the toy model where ARs can interact
only consecutively with some number of neighbors on each of the left
and right.  Nonetheless it is possible to remain well-mixed
statistically with \emph{aggregating} nearest-neighbor interactions
\footnote{A similar phenomenon is present in Burgers' equation with
  L\'evy random initial data
  \cite{Bertoin98,MenonPego07}.}\nocite{Bertoin98,MenonPego07}: in the
case of the toy model it can be shown that the partitions $\Pi(\tau)$,
$\tau = 0,1,2,\ldots$, are \emph{exchangeable} in the sense that the
vector of lengths
\begin{equation}
  (b - a + 1: [a,b] \in \Pi(\tau))
\end{equation}
has a distribution which is invariant under permutations.

\paragraph{Observables and moments ---}

We present an example to show how this connection between depinning
and coagulation can be exploited: certain observables are immediately
related to the explicitly calculable moments of the solution to the
Smoluchowski equation.  Namely, supposing that the avalanche
triggering sites and stop sites inside ARs are uniformly distributed,
which is consistent with numerics for large ARs, we find expected
length $\xi$ and size $\mathcal{S}$ of an avalanche as
\begin{equation}
  \label{eq:xi-s-moments}
  \xi = \frac{2}{3} \frac{M_2}{M_1} + \frac{1}{2} \frac{M_1}{M_0} %
  \quad \text{and} \quad %
  \mathcal{S} = \frac{1}{12} \frac{M_3}{M_1} + \frac{1}{6}
  \frac{M_2}{M_0},
\end{equation}
respectively.  The blue curves in Figure~\ref{fig:plot-av-ar} (left)
plot the moment relations for $\mathcal{S}$ and $\xi$ from
\eqref{eq:xi-s-moments} using the statistics of the AR lengths
obtained from our simulations.  The black curve in the main panel is
obtained by evaluating the moments $M_i$ using the exact solution
\eqref{eq:golovin}.  The agreement with simulations over the scaling
regime is quite good, deteriorating close to threshold for the result
based on the exact Smoluchowski solution.  The main reason for this
discrepancy is that the finite model admits clusters only as large as
$L$, whereas no such restriction exists for the Smoluchowski equation.
Using \eqref{eq:golovin} and \eqref{eq:xi-s-moments} in the scaling
region, it is readily shown that $\mathcal{S} = \frac{9}{16} \xi^2$,
which implies the scaling relation $\gamma = 2\nu$; \emph{cf}.\
\eqref{eq:scalinglaw}.

\paragraph{Conclusion ---}

We have presented numerical evidence connecting depinning phenomena
with coagulation, and finish with several reasons this relationship
deserves further exploration.  First, the toy model discussed in this
letter is sufficiently tractable that we expect further analytical
results should be attainable.  For instance it may be possible to
explicitly relate the various time scales $\tau$ and $X$ for the toy
model and $t$ for the Smoluchowski equation, which would provide not
only relations between $\xi$ and $\mathcal{S}$, as presented above,
but also express these as functions of time.  Second, though the model
we discuss is considerably simplified, the essential
features---aggregation of depinned segments (our ARs), avalanches
which relieve load in a few localized interior areas (our stop sites)
while increasing it at the boundaries---seem to be applicable to a
broader class of pinning models.  On a macroscopic level, depinning in
these models would be expected to be still governed by a similar
coagulation process.  Third, combining the Brownian scaling limit
result for the threshold configuration of \cite{KM14} with the
observations in this letter may lead to a stochastic process
describing the macroscopic limit of depinning in these models.  The
second and third points can provide an explanation for the emergence
of universal features in such transitions.

\begin{acknowledgments}
  The authors would like to thank M.\ M.\ Terzi for useful
  discussions. MM also acknowledges discussions with \mbox{A.\ Bovier}
  during the initial phase of this work.
\end{acknowledgments}

\bibliography{smoluchowski-sandpile}

\end{document}